\documentclass[%
 reprint,
 amsmath,amssymb,
 aps,
]{revtex4-2}

\usepackage{graphicx}
\usepackage{dcolumn}
\usepackage{bm}
\usepackage{amsmath}
\newcommand{\K}{\mathbf{k}}
\newcommand{\V}{\mathbf{v}}
\DeclareMathOperator{\tr}{Tr}
\usepackage{CJKutf8}
\usepackage{appendix}

\begin{document}
\begin{CJK}{UTF8}{gbsn}

\title{Lorentz Transformation of the Energy Spectrum of the Equilibrium State of Massive Free Fields}

\author{Ruohan Xu (徐若涵)}
\thanks{Equal contribution}

\affiliation{%
 School of Physics, Peking University, Beijing, 100871, China\\
}%

\author{Tingzhang Shi (石霆章)}
\thanks{Equal contribution}

\affiliation{%
 School of Physics, Peking University, Beijing, 100871, China\\
}%

\author{H. T. Quan (全海涛)}\email{htquan@pku.edu.cn}
\affiliation{
 School of Physics, Peking University, Beijing, 100871, China
}
\affiliation{
 Collaborative Innovation Center of Quantum Matter, Beijing 100871, China
}
\affiliation{Frontiers Science Center for Nano-optoelectronics, Peking University, Beijing, 100871, China}

\date{\today}

\begin{abstract}
In previous studies of relativistic thermodynamics, the temperature of a static system, as perceived by a moving observer, has traditionally been treated as a scalar. 
This assumption has also been extended to the research on the cosmic microwave background. 
However, the validity of this assumption is a consequence of the massless nature of photons. 
More generally, when an observer is in relative motion to a system, the thermal equilibrium state is characterized by a four-vector temperature. 
In this paper, we study the non-interacting massive Bosonic and Fermionic field systems.
We derive the Lorentz transformation of the energy spectral density in the equilibrium state of these fields. 
In the massless limit for bosonic field, our results recover the transformation of black body radiation [G. W. Ford and R. F. O’Connell., Phys. Rev. E, \textbf{88},044101(2013)], which corresponds to a scalar temperature with dipole anisotropy.
For the massive fields, the moving equilibrium state cannot be characterized by a corresponding scalar temperature.
This result shows the necessity of introducing four-vector temperature in relativistic thermodynamics.
\end{abstract}

\maketitle
\end{CJK}
\section{Introduction}
Constructing a covariant theory of thermodynamics has been a long-standing problem \cite{hanggi09, moya17}.
The covariant definition of temperature is one of the most fundamental issues in relativistic thermodynamics.
Several transformations of temperature have been proposed \cite{plank08, einstein07, ott63, landsberg66, kampen68}.
Considering a system at the temperature $T$, and a moving observer whose velocity relative to the system is $\V$.
In 1907 and 1908, Plank and Einstein proposed that the temperature of the system perceived in the moving observer's frame is $T'=T/\gamma$, here $\gamma=1/\sqrt{1-\V^2}$ is the Lorentz factor \cite{plank08, einstein07} (we have chosen the natural unit $\hbar=c=1$ hereafter).
Half a century latter, in 1963, Ott suggested that the temperature perceived in the moving frame of reference should be $T'=\gamma T$ instead \cite{ott63}.
Shortly after Ott's proposal, Landsberg reviewed the previous work and recommended that the temperature was a Lorentz invariant, $T'=T$ \cite{landsberg66}. 

Beside the theoretical explorations, the discovery of the cosmic microwave background (CMB) provided an excellent platform for the experimental examination of the theories.
Shortly after Penzias and Wilson's observation of the CMB in 1964 \cite{penzias65}, a dipole anisotropy in CMB temperature was observed in the 1970s \cite{conklin69,smooth77}, and the energy spectrum satisfies a perfect black body spectrum (Plank's law).
However, the dipole anisotropy in temperature was not predicted by any proposal of scalar temperature transformation \cite{plank08, einstein07, ott63, landsberg66}.
The expression of the transformed temperature $T'$ only contains a factor $\gamma$, which was not related to the direction of the relative motion, and thus cannot explain the dipole anisotropy in CMB temperature.

The failure of the previous proposals of temperature transformations in describing the dipole anisotropy in the CMB temperature lies in the notion of equilibrium state.
When promoting theories of thermodynamics to a covariant form, i.e., invariant for arbitrary inertial observer, we need to extend the definition of equilibrium state to a general form \cite{nakamura09, kamran19, jihui23}.
In the special theory of relativity, the energy and three-momentum together form the four-momentum, and mix when undergoing Lorentz transformation.
In thermodynamics, the inverse temperature is conjugate to energy, therefore the covariant form of the inverse temperature should be a four-vector, which includes some components being conjugate to the three-momentum.
In 1968, van Kampen had suggested that the temperature should be defined as a Lorentz vector \cite{kampen68}.
The Lorentz transformation depends on the direction of the relative motion, therefore the four-vector tmeperature could be used to explain the experimental observations \cite{nakamura09, jihui23, kamran19}.

The Lorentz transformation of black body radiation has been well studied, and the energy spectrum of a moving equilibrium electromagnetic field has been derived \cite{ford13}.
The scalar temperature with dipole anisotropy has been predicted theoretically \cite{ford13,heer68, conklin68, henry68, peebles68}. 
However, we find the reduction of the four-vector temperature to a scalar temperature is a consequence of the massless nature of photons.
To demonstrate this result, we calculate the Lorentz transformation of massive equilibrium fields, not only for the photons, but also for massive free bosonic or fermionic fields.
In the massless limit for bosonic fields, our results reproduce the main results in Ref. \cite{ford13}.
We demonstrate that for massive field, the equilibrium spectrum cannot be described by a scalar temperature with dipole anisotropy anymore.
Instead, a four-vector temperature should be used to characterize the equilibrium state perceived in the moving observer's frame.
Moreover, the fermionic case of our result may be applied to the study of the cosmic neutrino background ($\mathrm{C\nu B}$) \cite{scott24}.

The remaining of this article is organized as follows:
In section \ref{LT field}, we use the Lorentz transformation of the equilibrium state of a massive free field to calculate the energy spectrum of the field perceived in arbitrary inertial frame, and explain the results under the framework of the four-vector temperature.
In section. \ref{properties}, we analyze some properties of the four-vector temperature, and demonstrate that for the massive field, the moving equilibrium state cannot be characterized by a scalar temperature with dipole anisotropy any longer.
In section. \ref{summary}, we summarize the main results.

\section{Equilibrium spectrum of massive field perceived in arbitrary inertial frame of reference}\label{LT field}

\subsection{Setup}
Consider a massive real scalar field $\phi$ with proper mass $m$ and Lagrangian density $\mathcal{L}$:
\begin{equation}
    \mathcal{L}= \frac{1}{2} \partial_\mu \phi \partial^\mu \phi +\frac{m^2}{2} \phi^2.
\end{equation}
Here we use the metric tensor $\eta^{\mu\nu}=\mathrm{diag}(1,-1,-1,-1)$.
The Hamiltonian $H$ of this field is:
\begin{equation}
    H = \int \frac{d^3\K}{(2\pi)^3 2\omega_\K} \frac{\omega_\K}{2} (a_k a_k^\dagger+a_k^\dagger a_k),
\end{equation}
here $\K$ denotes the three-wave vector, $\omega_\K$ is the angular frequency of the particle with wave-vector $\K$, which equals to $\sqrt{\K^2+m^2}$ due to the on-shell condition. Since on-shell 4-momentum $k=(\omega_\K,\K)$ has a one-to-one mapping to $\K$, we can also denote $\omega_\K$ as $\omega_k$ without ambiguity. $a_k$ and $a_k^{\dagger}$ are the annihilation and creation operator of a particle with four-momentum $k$ corresponding to the scalar field $\phi$.

We may define $ D k=\frac{d^3\K}{(2\pi)^3 2\omega_\K} $ as a (on-shell) Lorentz invariant integral measure: 
\begin{equation*}
     \int D(\Lambda k) f(\Lambda k) = \int D k f(k),
\end{equation*}
for any Lorentz transformation $\Lambda$ and any function $f$. Moreover, we can define a Lorentz invariant delta distribution: 
\begin{equation}
    \delta^*(k-k'):= (2\pi)^3 2 \omega_k \delta^3 (\K-\K' ),
\end{equation}
which satisfies $\delta^*(\Lambda k)=\delta^*(k)$.
The commutator of the creation and annihilation operators can then be expressed in $\delta^*$:
\begin{equation}
    [a_k,a_{k'}^\dagger]=\delta^*(k-k').
\end{equation}
The field system is in equilibrium state and at rest in the frame of reference $i$, and there is a moving observer in the frame of reference $i'$, which moves at a velocity $\V $ with respect to frame $i$.

\subsection{Spectral density at equilibrium }
Suppose a system is in a thermal state at a finite temperature $\beta$ and chemical potential $\tilde{\mu}$ in the frame $i$, the particle number operator is $N$.
If we denote $\alpha=\beta \tilde{\mu}$, the density operator of the thermal equilibrium state is $\rho_i= \frac{e^{-\beta H+\alpha N}}{\tr(e^{-\beta H+\alpha N})}$. 
Then the expectation value of an operator $A$ is $\langle A \rangle_i =\tr(\rho_i A)$.
The density operator and the expectation value are not Lorentz invariant. 
In frame $i$, we can derive the following result \cite{ford13}:
\begin{equation}\label{thermal state expectation}
    \begin{gathered}
        \langle a_k a_{k'}^\dagger + a_{k}^\dagger a_{k'} \rangle_i= \delta^*(k-k')\coth(\frac{\beta\omega_k-\alpha}{2}),\\
    \langle a_k a_{k'}\rangle_i =\langle a_{k}^\dagger a_{k'}^{\dagger} \rangle_i= 0.
    \end{gathered}
\end{equation}
Now we calculate the the energy density $\rho_E$:
\begin{equation}
    \begin{split}
        \rho_E =&\langle T^{00} \rangle_i \\
        =& \frac{1}{V} \langle H \rangle_i \\
        =& \frac{1}{V} \int D k \frac{\omega_k}{2}  \langle a_k a_{k}^\dagger + a_{k}^\dagger a_k \rangle_i
    \end{split}
\end{equation}
Here $T^{\mu \nu}$ represents the energy-momentum tensor, thus $T^{00}$ is the energy density.
Using Eq. (\ref{thermal state expectation}) and the box-normalization relation $\delta^3(0)=\frac{V}{(2\pi)^3}$, we obtain
\begin{equation}
    \begin{split}
        \rho_E =& \int \frac{d^3\K}{(2 \pi)^3} \frac{\omega_k}{2} \coth(\frac{\beta \omega_k-\alpha}{2}) \\
        =& \int |\K|^2 d\Omega d|\K| \frac{\omega_k}{2(2\pi)^3} \coth(\frac{\beta \omega_{\K}-\alpha}{2}),
    \end{split}
\end{equation}
here $d\Omega$ is the solid angle of the wave vector $\K$.
Now we define the spectral density $ \rho(\omega,\hat{\K}) $ to satisfy the relation $ \rho_E= \int \rho(\omega,\hat{\K}) d\omega d\Omega $, where $\omega=\omega_{\K}=\sqrt{\K^2+m^2}$ and $\hat{\K}$ is the unit vector along $\K$. 
For the massless case we have $\omega=|\K|$, then 
\begin{equation}
    \rho(\omega,\hat{\K}) = \frac{\omega^3  }{2(2\pi)^3} \coth(\frac{\beta \omega-\alpha}{2}).
\end{equation}
For black body radiation, $\tilde{\mu}=0, \alpha=0$, the result is just Planck's law (including zero-point energy) without the 2-fold spin degeneracy.
If $m\neq 0$, then $d|\K|= \frac{\omega}{\sqrt{\omega^2-m^2}}d\omega$, and we can see \cite{pathria4th}
\begin{equation}
    \rho(\omega,\hat{\K}) = \frac{\omega^2 \sqrt{\omega^2-m^2}  }{2(2\pi)^3} \coth(\frac{\beta \omega-\alpha}{2}).
\end{equation}
The spectral density is isotropic, because the observer is at rest with the equilibrium field.

\subsection{Spectral density perceived in the moving observer's frame of reference}
Consider an observable $A$ in frame of reference $i'$ with a 3-velocity $\V$.  
The Lorentz transformation from frame $i$ to frame $i'$ is denoted as $\Lambda$.
The density operator of the thermal state in frame $i'$ is $\rho_{i'}$.
Then the expectation value of $A$ in frame $i'$ is $\langle A \rangle_{i'}= \tr(\rho_{i'} A)$, where $\rho_{i'}= U(\Lambda) \rho_i U(\Lambda)^\dagger$, and $U(\Lambda)$ is the unitary operator corresponding to the Lorentz transformation $\Lambda$. 
This state can be viewed as a generalized thermal state: $\rho_{i'}=\frac{e^{-\beta u^\mu P_\mu+\alpha N}}{\tr(e^{-\beta u^\mu P_\mu+\alpha N})}$, here $u_\mu$ is the four-velocity of frame $i$ relative to frame $i'$. 
With the cyclic property of trace, we have
\begin{equation}
    \begin{split}
    \langle A \rangle_{i'}=& \tr (U(\Lambda) \rho_i U(\Lambda)^\dagger A)\\
    =& \tr ( \rho_i U(\Lambda)^\dagger A U(\Lambda))\\
    =& \langle U(\Lambda)^\dagger A U(\Lambda) \rangle_i .
    \end{split}
\end{equation}


Using this relation, the energy density is 
\begin{equation}
    \begin{split}
        \rho_E'=& \frac{1}{V'} \int D k' \frac{\omega_{k'}}{2}  \langle a_{k'} a_{k'}^\dagger + a_{k'}^\dagger a_{k'} \rangle_{i'}\\
        =& \frac{1}{V'} \int D k' \frac{\omega_{k'}}{2}  \langle a_{\Lambda^{-1}k'} a_{\Lambda^{-1}k'}^\dagger + a_{\Lambda^{-1}k'}^\dagger a_{\Lambda^{-1}k'} \rangle_i
    \end{split}
\end{equation}
By using Eq. (\ref{thermal state expectation}) and the invariance of $\delta^*$, we have
\begin{equation}
    \begin{split}
        &\rho_E'\\
        =& \frac{1}{V'} \int D k' \frac{\omega_{k'}}{2}  [\delta^*(\Lambda^{-1}k'-\Lambda^{-1}k')\coth(\frac{\beta \omega_{\Lambda^{-1}k'}-\alpha}{2} )]\\
        =& \frac{1}{V'} \int D k' \frac{\omega_{k'}}{2}  [\delta^*(k'-k')\coth(\frac{\beta \omega_{\Lambda^{-1}k'}-\alpha}{2} )]\\
        =& \frac{1}{V'} \int D k' \frac{\omega_{k'}}{2} (2\pi)^3 2\omega_{k'} \delta^3(\K'-\K')\coth(\frac{\beta \omega_{\Lambda^{-1}k'}-\alpha}{2} ).
    \end{split}
\end{equation}
It is clear the only non-covariant term is $\coth(\frac{\beta \omega_{\Lambda^{-1}k'}-\alpha}{2} )$, which is the effect of finite temperature.  Again by using $\delta^3(\K'-\K')=\frac{V'}{(2\pi)^3}$, we have
\begin{equation}
\rho_E'=\int D k' \omega_{k'}^2 \coth(\frac{\beta \omega_{\Lambda^{-1}k'}-\alpha}{2} ),
\end{equation}
so the spectral distribution in frame $i'$ is 
\begin{equation}\label{spectral density in moving frame}
\rho'(\omega',\hat{\K}')= \frac{|\K'(\omega')|^2 \omega' }{2(2\pi)^3} \frac{d |\K'|}{d\omega'} \coth(\frac{\omega_{\Lambda^{-1}k'} -\alpha}{2} ).
\end{equation}

This result actually contains the key result in Ref. \cite{ford13} as a massless limit. If we take $|\K'|=\omega'$ and chemical potential $\tilde{\mu}=0$, then $\alpha=0$, and Eq. (\ref{spectral density in moving frame}) simplifies to 
\begin{equation}
    \rho'(\omega',\hat{\K}')= \frac{\omega'^3 }{2(2\pi)^3}  \coth(\frac{\beta \gamma (1 + \hat{\K}'\cdot \V)\omega'}{2}),
\end{equation}
which, if multiplied by a 2-fold spin degeneracy, is just the main result in Ref. \cite{ford13}. In this case, in any direction $\hat{\K}'$, we can see the energy spectrum over $\omega'$ is a perfect black body spectrum with a modified temperature. In the more general case $m \neq 0$, we have
\begin{equation}
    \begin{split}
         \rho'(\omega',\hat{\K}') =& g_s \frac{\omega'^2 \sqrt{\omega'^2-m^2}  }{2(2\pi)^3}\\
         \cdot & \coth(\frac{\beta \gamma(\omega' + \hat{\K}'\cdot \V \sqrt{\omega'^2-m^2} )-\alpha}{2}) .
    \end{split}
\end{equation}
For fermions, we can obtain similarly  
\begin{equation}
    \begin{split}
        \rho'(\omega',\hat{\K}') = &- g_s\frac{\omega'^2 \sqrt{\omega'^2-m^2}  }{2(2\pi)^3} \\
        &\cdot \tanh(\frac{\beta \gamma(\omega' + \hat{\K}'\cdot \V \sqrt{\omega'^2-m^2} )-\alpha}{2}) .
    \end{split}
\end{equation}
Here $g_s$ account for the spin degeneracy (for details, please see Appendix \ref{app:fermion}).

\subsection{Equilibrium state characterized by four-vector temperature}
For a system with the inverse temperature $\beta$ defined in the rest frame, and with a four-velocity $u^\mu$ measured in the moving observer's frame of reference, the four-vector temperature $\beta^\mu$  of the system is defined as \cite{kampen68}
\begin{equation}
    \beta^\mu=\beta u^\mu.
\end{equation}
We now show that the spectral density should be written as a function of  $\beta^\mu$ instead of scalar temperature $\beta$.

In the rest frame of the system, the four-velocity of the system itself is $u^\mu=(1,0,0,0)$, thus the four-vector temperature is simply $\beta^\mu=(\beta,0,0,0)$.
In the moving frame $i'$, the four-velocity of the system is $u'^\mu=\gamma(1,-\V)$, and the contravariant four-vector temperature becomes $\beta'^\mu=\gamma(\beta,-\beta v_x,-\beta v_y,-\beta v_z)$. Correspondingly, the covariant four-vector temperature is $\beta'_\mu=\eta_{\mu\nu}\beta'^{\nu}=\gamma(\beta,\beta v_x,\beta v_y,\beta v_z)$.

The spectral density of energy is then fully characterized by the four-vector temperature of the field and the four-momentum of the particles.
For bosons, 
\begin{equation}\label{transformed spectrum}
    \begin{split}
        &\rho'(\omega',\hat{\K}') \\
        =&  g_s \frac{\omega'^2 \sqrt{\omega'^2-m^2}  }{2(2\pi)^3} \coth(\frac{\beta \gamma(\omega' + \V \cdot \hat{\K}'\sqrt{\omega'^2-m^2} )-\alpha}{2})\\
        =& g_s\frac{\omega'^2 \sqrt{\omega'^2-m^2}  }{2(2\pi)^3} \coth(\frac{\beta'_\mu P'^\mu-\alpha}{2}),
    \end{split}
\end{equation}
here $P'^\mu =(\omega',\hat{\K}'\sqrt{\omega'^2-m^2})=(\omega',\hat{\K}'|\K'|)=(\omega',\K')$ is the four-momentum of the particles.

In the rest frame of the field (frame $i$), the only non-zero component in the four-vector temperature is $\beta_0=\beta$, therefore $\beta_\mu P^\mu=\beta_0 \cdot \omega+ 0 \cdot P_x + 0 \cdot P_y + 0 \cdot P_z = \beta \omega$, and the three-momentum does not appear in the spectral density.
However, the spectrum characterized by four-vector temperature is universal in any inertial frame. 
Generally, we can remove the notation $'$ in Eq. (\ref{transformed spectrum}), and the expression is valid in arbitrary inertial frame of reference.
For bosons, the covariant spectral density of the generalized thermal state $\rho=\frac{e^{-\beta u^\mu P_\mu+\alpha N}}{\tr(e^{-\beta u^\mu P_\mu+\alpha N})}$ is
\begin{equation}\label{bosons density}
    \rho(\omega,\hat{\K}) =  g_s  \frac{\omega^2 \sqrt{\omega^2-m^2}  }{2(2\pi)^3} \coth(\frac{\beta_\mu P^\mu -\alpha}{2});
\end{equation}
and for fermions, the spectral density is 
\begin{equation}\label{fermions density}
    \rho(\omega,\hat{\K}) = - g_s \frac{\omega^2 \sqrt{\omega^2-m^2}  }{2(2\pi)^3} \tanh(\frac{\beta_\mu P^\mu -\alpha}{2}),
\end{equation}
 $\beta_\mu$ is the covariant four-vector temperature of the field, and $P^\mu$ is the contravariant four-momentum of a particle with energy $\omega$ and moves in the direction $\hat{\K}$.
 $\beta_\mu$ and $P^\mu$ can be defined in any inertial frame and satisfy the Lorentz transformation of four-vectors.
 $\alpha$ corresponds to the chemical potential of a particle.
 Since the particle number is a Lorentz scalar, the corresponding Lagrange multiplier $\alpha$, which represents the conservation of particle number, should also be a Lorentz scalar.

 The $\mathrm{C\nu B}$ is supposed to be an equilibrium field of neutrinos, which could be treated as a free field.
 The rest mass of the neutrino is estimated to be $10^{-2}eV \sim 10^{-1}eV$ \cite{scott24}.
 The temperature of $\mathrm{C\nu B}$ is predicted to be around $1.95K$, corresponding to an energy scale of $10^{-4}eV$ \cite{pathria4th, scott24}.
 The thermal fluctuation is comparable to the rest energy of neutrinos.
 Although the spectrum of $\mathrm{C\nu B}$ may not be ultra-relativistic for some flavours of neutrinos, it may still be influenced by relativistic effects.
 Generally, the $\mathrm{C\nu B}$ should be characterized by a relativistic spectrum.

\section{Analysis of the properties of the four-vector temperature}\label{properties}


From the study of the equilibrium state of the free fields, we observe a distinct difference between the massive and massless systems:
For massless particles like photons, the equilibrium state distribution can be characterized by a scalar temperature with dipole anisotropy.
However, for the massive particles whose momentum are no longer proportional to the energy, it is impossible to characterize the equilibrium state perceived in the moving observer's frame with a scalar temperature, and it is necessary to express the equilibrium state distribution in the form of four-vector temperature and four-momentum.

\subsection{Failure of scalar temperature description for massive fields}
For massless particles like photons, the spectrum distribution in a specific direction $\hat{\K'}$ satisfies the exact black-body radiation spectrum, even in a moving frame of reference (like frame $i'$ which has a velocity $\V$ relative to frame $i$).
The spectral density in direction $\hat{\K'}$ is $\rho(\omega',\K')=(\omega'/2\pi)^3\coth[\gamma (1+\V \cdot \hat{\K'})\omega'/2T]$, which corresponds to a black body spectrum at rest with an effective temperature $T'(\hat{\K'})$,
\begin{equation}
    T'(\hat{\K}')=\frac{T}{\gamma (1+\V \cdot \hat{\K}')}=\frac{T}{\gamma(1+ |\V|\cos{\theta})},
\end{equation}
here $\theta$ is the angle between the direction of observation $\hat{\K}'$ and the direction of the relative motion $\hat{\V}$.
This result has been studied theoretically \cite{ford13, kamran19, nakamura09,conklin68, henry68, peebles68} and observed experimentally \cite{conklin69,smooth77,cobeweb}.

This correspondence is a consequence of the massless nature of photons.
In Eq. (\ref{bosons density}), for a photon whose $m=0$, the absolute value of three momentum $|\K'|$ is proportional to its energy $\omega'$, therefore the contraction $\beta'_\mu P'^\mu \propto \omega'$, and the ratio $\omega'/\beta'_\mu P'^\mu$ could be regarded as an effective temperature $T'_{eff}=\omega'/\beta'_\mu P'^\mu=T/\gamma(1+\V \cdot \hat{\K'})$, which is independent of $\omega'$.

However, for massive particles, the corresponding ratio is $\omega'/{\beta'_\mu P'^\mu}=T/{\gamma(1+\V \cdot \hat{\K}'\sqrt{1-(\frac{m}{\omega'})^2})}$, which is a function of both $\hat{\K}'$ and $\omega'$.
This expression contains $\omega'$, thus $\omega'/\beta'_\mu P'^\mu$ cannot be regarded as an effective temperature anymore for massive fields. 
This explains why we can no longer use a scalar temperature to characterize the equilibrium state of massive fields.

Here we can see that it is just a coincidence that the spectrum of the CMB measured by a moving observer (the observer fixed on earth) can be characterized by a scalar temperature, because only for massless particles, the four-vector temperature can be reduced to a scalar temperature with dipole anisotropy.
Generally, the moving observer needs the four-vector temperature to characterize the equilibrium state in the rest frame.

\subsection{Some explicit illustrations}
Let us take the spectral density of fermions as an example, and recall the results in Eq. (\ref{fermions density}).
We would like to highlight two situations: the zero-temperature and infinite-temperature cases.

The first illustration is the vacuum state of the field, and the corresponding chemical potential is $\tilde{\mu}=0, \alpha=0$.
The factor $-\tanh(\beta_\mu P^\mu/2)=-\tanh(\beta (u_\mu P^\mu)/2)$, $\beta$ is the rest temperature of the equilibrium field, and $u_\mu P^\mu$ is the energy measured in the rest frame of the field, which is positive definite.

For the zero-temperature state, $\beta \to \infty$, and $-\tanh(\beta (u_\mu P^\mu)/2) \to -1$, which means that there is no particle excited in any mode.
For two fermionic field systems with different velocities and zero temperatures, the difference between $u_\mu$ is finite, and the difference between four-vector temperatures even diverges. 
However, there is no difference in the spectral density.
This result is intuitive: the vacuum state is invariant under Lorentz transformation.

In the infinite temperature state, $\beta =0, \alpha=0$, and $-\tanh(\beta (u_\mu P^\mu)/2) =0$, this temperature-dependent factor also becomes a constant, meaning that every state is equally occupied.
For two fermionic field systems with different velocities and infinite temperatures, the difference between $u_\mu$ is finite, but the difference between the four-vector temperatures is zero ($\beta = 0$, thus $\beta_\mu = \beta u_\mu = 0$ for arbitrary $u_\mu$, so there is no difference between $\beta_\mu$ for infinite temperature systems with different velocities).
This result is intuitive too: the equally occupied state should also be invariant under Lorentz transformation.

In the intermediate cases, when the temperatures are finite, the necessary and sufficient condition for two field systems, $S$ and $S'$ (with the same $m$), to have the same spectral density is
\begin{equation*}
    \begin{split}
         \beta_\mu =& \beta'_\mu, \\
          \tilde{\mu} =& \tilde{\mu'}.
    \end{split}
\end{equation*}
Here $\beta_\mu/\beta'_\mu$ and $\tilde{\mu}/\tilde{\mu'}$ represent the four-vector temperature and the chemical potential of $S/S'$.
Under this circumstance,
\begin{equation}
    \begin{split}
        \beta=\sqrt{\beta_\mu \beta^\mu} =& \sqrt{\beta'_\mu \beta'^\mu} = \beta',\\
        \alpha=\beta\tilde{\mu}=& \beta'\tilde{\mu'}=\alpha',\\
        u_\mu=\frac{\beta_\mu}{\beta} =& \frac{\beta'_\mu}{\beta'}=u'_\mu.
    \end{split}
\end{equation}
The equivalence of four-vector temperature implies that the two systems move at the same velocity (the same $u_\mu$), and have the same rest temperature in each's co-moving frame.
In this case, the thermal equilibrium relation becomes straightforward.

It is worthwhile to point out that a single Lorentz scalar $\beta$ is insufficient to characterize the thermal equilibrium relation.
We can easily construct two thermal equilibrium systems $R_1$ and $R_2$ with the same rest temperature $\beta$, the same chemical potential $\tilde{\mu}$, and a relative velocity $\V$. 
In the rest frame of $R_1$, the energy spectrum of $R_1$ and $R_2$ are
\begin{equation*}
    \begin{split}
        &\rho_1(\omega,\hat{\K})\\
        =&-g_s \frac{\omega^2\sqrt{\omega^2-m^2}}{2(2\pi)^3}\tanh(\frac{\beta \omega-\alpha}{2}),\\
        &\rho_2(\omega,\hat{\K})\\
        =& -g_s \frac{\omega^2\sqrt{\omega^2-m^2}}{2(2\pi)^3}\tanh(\frac{\gamma \beta (\omega+\V\cdot \hat{\K}\sqrt{\omega^2-m^2})-\alpha}{2}).
    \end{split}
\end{equation*}
If we let the two fields couple to each other and allow particle flux in a specific direction and a specific frequency (via a filter, collimator, or a resonant cavity), then the net energy flow is typically non-zero, indicating that the two field systems are not in equilibrium with each other.

The discussions above demonstrate that the spectral density characterized by four-vector temperature is sufficient to characterize the thermal equilibrium relation.
The four-vector temperature provides a covariant description of the thermal equilibrium states.

\subsection{Classical and non-relativistic limit of the spectrum}
In the classical limit (quantum statistical effects can be neglected), we expect that the bosonic spectrum recovers a J\"uttner-Synge distribution.
Furthermore, in the non-relativistic limit, we expect that the spectrum recovers a shifted Maxwellian distribution.

To obtain the classical limit, we should firstly subtract the influence of zero-point fluctuation.
Denote the zero-point spectral density as $\rho_0(\omega)$.
For the bosons,
\begin{equation}
    \begin{split}
        &\rho(\omega,\hat{\K})-\rho_0(\omega)\\
        =& g_s\frac{\omega^2\sqrt{\omega^2-m^2}}{(2\pi)^3}[\frac{1}{2}\coth(\frac{\gamma\beta (\omega+\V\cdot\K)-\alpha}{2})-\frac{1}{2}]\\
        =&g_s\frac{\omega^2\sqrt{\omega^2-m^2}}{(2\pi)^3}\frac{\exp(\frac{-\gamma\beta (\omega+\V\cdot\K)+\alpha}{2})}{2\sinh(\frac{\gamma\beta (\omega+\V\cdot\K)-\alpha}{2})}\\
        \approx& g_s\frac{\omega^2\sqrt{\omega^2-m^2}}{(2\pi)^3}\exp(-\gamma\beta (\omega+\V\cdot\K)+\alpha)\\
        =&g_s e^{\alpha}\frac{\omega^2\sqrt{\omega^2-m^2}}{(2\pi)^3}\exp(-\beta_\mu P^\mu).
    \end{split}
\end{equation}
Here we have used the approximation
\begin{equation*}
    2\sinh(\frac{\gamma\beta (\omega+\V\cdot\K)-\alpha}{2}) \approx \exp(\frac{\gamma\beta (\omega+\V\cdot\K)-\alpha}{2}),
\end{equation*}
because
\begin{equation*}
    \exp(\frac{-\gamma\beta (\omega+\V\cdot\K)+\alpha}{2})\approx 0.
\end{equation*}
This classical approximation requires that the density of particles is low enough so that we can ignore the exchange interactions of identical bosons.
In this limit, the J\"uttner-Synge distribution is recovered.

The non-relativistic condition requires that $|\V| \ll 1$, $|\K|\ll m$ and $\beta \gg 1/m$.
We can then perform the Taylor expansion
\begin{equation}
    \begin{split}
        \omega=&\sqrt{\K^2+m^2}\approx m+\frac{\K^2}{2m},\\
        \gamma=&\frac{1}{\sqrt{1-\V^2}}\approx 1+\frac{\V^2}{2}.\\
    \end{split}
\end{equation}
Notice that $\K=\hat{\K}\sqrt{\omega^2-m^2}$ is an exact result, we obtained
\begin{equation}
    \begin{split}
        \gamma\beta (\omega+\V\cdot\K) &\approx 
    \beta m+\beta(\frac{1}{2}m\V^2+ \frac{\K^2}{2m}+\V\cdot\K)\\
    &=\beta (m+\frac{(\K+m\V)^2}{2m}).
    \end{split}
\end{equation}
With both the classical limit and the non-relativistic expansion, we have
\begin{equation}
    \begin{split}
        & g_s\frac{\omega^2\sqrt{\omega^2-m^2}}{(2\pi)^3}\exp(-\gamma\beta (\omega+\V\cdot\K)+\alpha)\\
        \approx & g_s e^{-\beta m+\alpha}\frac{(m^2+\K^2)|\K|}{(2\pi)^3}e^{-\beta \frac{(\K+m\V)^2}{2m}}.
    \end{split}
\end{equation}
The relativistic chemical potential already includes the rest energy $mc^2$.
If we hope to recover the non-relativistic results, we may redefine the non-relativistic chemical potential as $\tilde{\mu}_{nr}=\tilde{\mu}-m$, then
\begin{equation*}
    e^{-\beta m+\alpha}=e^{\beta(\tilde{\mu}-m)}=e^{\beta \tilde{\mu}_{nr}}.
\end{equation*}
In classical mechanics, the spin degrees of freedom can be ignored so $g_s=1$, and the classical non-relativistic spectrum finally becomes
\begin{equation}
    \begin{split}
         &\rho(\omega,\hat{\K})-\rho_0(\omega)\\
         \approx & g_s e^{-\beta m+\alpha}\frac{(m^2+\K^2)|\K|}{(2\pi)^3}e^{-\beta \frac{(\K+m\V)^2}{2m}}\\
         =&\frac{(m^2+\K^2)|\K|}{(2\pi)^3}e^{\beta (\tilde{\mu}_{nr}- \frac{(\K+m\V)^2}{2m})},
    \end{split}
\end{equation}
which is exactly a shifted Maxwellian distribution with a center of mass velocity $-\V$.

For the fermions, we observe the shift of the Fermi surface in the zero-temperature and non-relativistic limit.
Denote the Heaviside step function as $\theta(x)$, then
\begin{equation}
    \begin{split}
        &\rho(\omega,\hat{\K})-\rho_0(\omega)\\
        =& g_s\frac{\omega^2\sqrt{\omega^2-m^2}}{(2\pi)^3}[-\frac{1}{2}\tanh(\frac{\gamma\beta (\omega+\V\cdot\K)-\alpha}{2})+\frac{1}{2}]\\
        \approx & g_s \frac{(m^2+\K^2)|\K|}{(2\pi)^3} \theta(\alpha-\gamma\beta (\omega+\V\cdot\K))\\
        \approx & g_s \frac{(m^2+\K^2)|\K|}{(2\pi)^3} \theta(\alpha-\beta m-\beta \frac{(\K+m\V)^2}{2m}).
    \end{split}
\end{equation}
Using the non-relativistic chemical potential $\tilde{\mu}_{nr}=\tilde{\mu}-m$, we express the step function as
\begin{equation}
    \begin{split}
        &\theta(\alpha-\beta m-\beta \frac{(\K+m\V)^2}{2m})\\
        =&\theta[\beta(\tilde{\mu}_{nr} - \frac{(\K+m\V)^2}{2m})]\\
        =&\theta(\tilde{\mu}_{nr} - \frac{(\K+m\V)^2}{2m}),
    \end{split}
\end{equation}
from which we found that the non-relativistic chemical potential in the low-temperature limit is exactly the Fermi energy $\varepsilon_F$, and the Fermi surface is shifted by a constant $-m\V$ in the momentum space due to the relative motion between the observer and the system.
The non-relativistic analysis proves that the four-vector temperature is consistent with the non-relativistic statistical mechanics.

\subsection{Equilibrium state with conserved charge}
The field system of interest may have some conserved charges. 
In this circumstance, the field will not thermalize to the full Hilbert space, but within a small section with a certain value of charge. 
In this case, we should introduce a generalized chemical potential $\tilde{\alpha}$ as a Lagrange multiplier for the charge restriction. 
For example, the complex scalar field $\phi$, with Lagrangian density
\begin{equation}
    \mathcal{L}= \partial_{\mu} \phi^\dagger \partial^{\mu} \phi + m^2 \phi^\dagger \phi,
\end{equation}
has a globally conserved $U(1)$ charge $ Q= \int d^3 x j^0 (x) $, where
\begin{equation}
    j^{\mu}= \mathrm{Im}(\phi^\dagger \partial^{\mu} \phi-\phi\partial^{\mu} \phi^\dagger),
\end{equation}
so we may refine our definition of thermal state to be a generalized Gibbs form:
\begin{equation}
    \rho(\beta^\mu,\tilde{\alpha})= \frac{e^{-\beta^\mu P_\mu-\tilde{\alpha} Q}}{tr(e^{-\beta^\mu P_\mu-\tilde{\alpha} Q})}.
\end{equation}
Obviously $Q$ is invariant under the action of Poincare group. 
So to calculate the energy spectral density, we simply repeat the previous process and the result for the complex scalar field is 
\begin{equation}
    \begin{split}
        \rho(\omega,\hat{\K}) =& \frac{\omega^2 \sqrt{\omega^2-m^2}  }{2(2\pi)^3} [\coth(\frac{\beta_\mu P^\mu+\tilde{\alpha}}{2})\\
        &+\coth(\frac{\beta_\mu P^\mu-\tilde{\alpha}}{2})].
    \end{split}
\end{equation}
The physics behind this result is quite clear: it's just the total energy of particles with chemical potential $\tilde{\alpha}/\beta$ and antiparticles with chemical potential $-\tilde{\alpha}/\beta$. This simple result is also comparable with the discussion of electron-positron equilibrium in Ref. \cite{pathria4th}. 

Particle number $N$ and four-momentum $P^\mu$ are just some of the conserved charges.
For systems with higher symmetry, we can simply introduce the corresponding Lagrange multipliers and express the thermal equilibrium state in the generalized Gibbs form.
It is worth pointing out that the four-vector temperature reflects the energy-momentum conservation of a relativistic system, which is valid whenever there is Poincare group symmetry. 
Therefore the four-vector temperature is always necessary in the relativistic thermodynamics.

\section{Summary}\label{summary}
In this paper, we drive the equilibrium spectrum distribution of either bosonic or fermionic massive free field perceived in a moving frame of reference. 
In the massless limit, our result recovers the main result of Ref.~\cite{ford13}, and is consistent with the observation \cite{conklin69,smooth77,cobeweb}.
However, our study are generalized to both bosonic and fermionic fields. 
Furthermore, the analytical expression of the spectrum proves that for a massive field system, an effective scalar temperature description is insufficient, and the perfect black-body spectral density of CMB is nothing more than a coincidence due to the massless nature of the photons.
Therefore, it is necessary to introduce the four-vector temperature to characterize the thermal equilibrium state, and the spectral density based on the four-vector temperature is sufficient to provide a covariant description of the equilibrium state perceived from a moving frame of reference.
Our results might be applied to realistic systems like the $\mathrm{C\nu B}$ \cite{scott24}, which is typically considered as a relativistic massive free field.
If the rest mass of neutrinos cannot be neglected, the spectrum of $\mathrm{C\nu B}$ would be quite different from CMB's: the spectrum would not be characterized by any scalar temperature with dipole anisotropy, and the four-vector temperature description of the $\mathrm{C\nu B}$ spectrum would be a necessity.

\appendix

\section{Calculation of the fermionic thermal spectrum}\label{app:fermion}
Here we consider a Majorana field $\psi$ with Lagrangian density
\begin{equation}
    \mathcal{L}= \bar{\psi} (i\gamma^\mu \partial_\mu -m) \psi 
\end{equation}
and Hamiltonian
\begin{equation}
    H = \sum_s \int \frac{d^3\K}{(2\pi)^3 2\omega_\K} \frac{\omega_\K}{2} (-b_{ks} b_{ks}^\dagger+b_{ks}^\dagger b_{ks}).
\end{equation}
Here $\gamma^\mu$ are Dirac matrices, $b_{ks}$ and $b_{ks}^\dagger$ denote the fermionic creation and annihilation operators for a particle with four-momentum $k$ and spin $s$. The anti-commutation relation is $\{b_{ks},b_{k's'}^\dagger\}=(2\pi)^3 2\omega_k \delta (k-k') \delta_{ss'} $.

If we still define the density operator of the thermal state in frame $i$ to be: $\rho_{i}=\frac{e^{-\beta H+\alpha N}}{\tr(e^{-\beta H+\alpha N})}$, then the expectation values of the following operators in frame $i$ are
\begin{equation}
    \begin{gathered}
        \langle b_{ks} b_{k's'}^\dagger + b_{ks}^\dagger b_{k's'} \rangle_i=- \delta^*(k-k')\tanh(\frac{\beta\omega_k-\alpha}{2})\delta_{ss'},\\
    \langle b_{ks} b_{k's'}\rangle_i =\langle b_{ks}^\dagger b_{k's'}^{\dagger} \rangle_i= 0.
    \end{gathered}
\end{equation}

Correspondingly, if we perform the Lorentz transformation from frame $i$ to frame $i'$, then the energy density of the field should be
\begin{equation}
    \begin{split}
        \rho_E'=&-\sum_s \frac{1}{V'} \int D k' \frac{\omega_{k'}}{2}  [\delta^*(\Lambda^{-1}k'-\Lambda^{-1}k')\tanh(\frac{\beta \omega_{\Lambda^{-1}k'}-\alpha}{2} )]\\
        =& -g_s\frac{1}{V'} \int D k' \frac{\omega_{k'}}{2}  [\delta^*(k'-k')\tanh(\frac{\beta \omega_{\Lambda^{-1}k'}-\alpha}{2} )]\\
        =& -g_s\frac{1}{V'} \int D k' \frac{\omega_{k'}}{2} (2\pi)^3 2\omega_{k'} \delta^3(\K'-\K')\tanh(\frac{\beta \omega_{\Lambda^{-1}k'}-\alpha}{2} ),
    \end{split}
\end{equation}
so the spectral distribution in frame $i'$ is 
\begin{equation}
    \begin{split}
        \rho'(\omega',\hat{\K}') = &- g_s\frac{\omega'^2 \sqrt{\omega'^2-m^2}  }{2(2\pi)^3} \\
        &\cdot \tanh(\frac{\beta \gamma(\omega' + \hat{\K}'\cdot \V \sqrt{\omega'^2-m^2} )-\alpha}{2}) ,
    \end{split}
\end{equation}
here $g_s$ account for the spin degeneracy.

\bibliography{apssamp}

\end{document}